\documentclass[11pt,twocolumn]{article}
\usepackage{arxiv}
\usepackage{float}
\usepackage[utf8]{inputenc} 
\usepackage[T1]{fontenc}    
\usepackage{hyperref}       
\usepackage{url}            
\usepackage{booktabs}       
\usepackage{amsfonts}       
\usepackage{nicefrac}       
\usepackage{microtype}      
\usepackage{lipsum}
\usepackage{fancyhdr}       
\usepackage{graphicx}       

\graphicspath{{media/}}     

\AtBeginDocument{%
\providecommand\BibTeX{{%
\normalfont B\kern-0.5em{\scshape i\kern-0.25em b}\kern-0.8em\TeX}}}

\usepackage[backend=bibtex,
style=numeric]{biblatex} 

\begin{document}

\title{Knight Watch – Ubiquitous Computing Enhancements to Sleep Quality with
    Acoustic Analysis}


\author{
    Andrew Ajemian\\
    Department of Computer Science\\
    Georgia Institute of Technology\\
    ajemian@gatech.edu \\
  \and
    John Knight\\
    Department of Computer Science\\
    Georgia Institute of Technology\\
    jknight78@gatech.edu \\
  \and
    Tommy Nguyen\\
    Department of Computer Science\\
    Georgia Institute of Technology\\
    tnguyen648@gatech.edu \\
  \and
    John O’Connor\\
    Department of Computer Science\\
    Georgia Institute of Technology\\
    john.oconnor@gatech.edu \\
}

\date{}

\twocolumn[
    \maketitle
]


\begin{abstract}
This project introduces a wearable, non-intrusive device for snoring detection
and remediation, designed to be placed under or alongside a pillow. The device
uses sensors and machine learning algorithms to detect snoring and employs
gentle vibrations to prompt positional changes, thereby reducing snoring
episodes. The device is capable of connecting via an API to a cloud-based
platform for the analysis of snoring sleep patterns and environmental context.
The paper details the development from concept to prototype, emphasizing the
technical challenges, solutions, and alignment with ubiquitous computing in
sleep quality improvement.
\end{abstract}

\keywords{Snoring; Obstructive Sleep Apnea; Prophylaxis; Machine Learning;
Wearable Computing; Ubiquitous Computing; Haptic Feedback; Cloud-based Analysis}


\section{Introduction}

Snoring, often trivialized, is a prevalent sleep disturbance with far-reaching
effects on health and well-being. This project seeks to address snoring with an
innovative solution: a wearable, non-intrusive device designed to be placed
under or alongside a pillow, capable of detecting and mitigating snoring
episodes. Snoring is not only a common annoyance affecting sleep quality but is
also linked to more severe health problems: including obstructive sleep apnea
(OSA) and cardiovascular issues \cite{sleep_disorder}.

The motivation for our approach stems from the limitations of current snoring
interventions, which range from invasive surgical procedures to cumbersome
Continuous Positive Airway Pressure (CPAP) machines \cite{surgical_snoring}.
These methods, although effective, have low user compliance due to their
intrusive nature. Our solution, leveraging advancements in sensor technology,
machine learning, and ubiquitous computing, aims to offer a discreet,
user-friendly alternative.

This paper outlines the development of our prototype ubiquitous computing
device, from the conceptualization of using tactile feedback for snoring
remediation to the implementation of cloud-based analytics by a collaborating
Health Informatics team for sleep pattern analysis. We grounded our approach in
the state-of-the-art, drawing from research in sleep science, wearable
technology, and machine learning. The team’s goal is to not only provide a
practical solution to snoring but also contribute to sleep health research.

\section{BACKGROUND}

In recent years, sleep research has seen a growing interest in leveraging
technological advancements to address sleep disorders \cite{consumer_tech}.
Studies have explored acoustic analysis, wearable sensors, and haptic feedback
in monitoring and mitigating snoring. Machine learning models show promise in
distinguishing snoring sounds from other nocturnal noises, leading to more
accurate detection and intervention \cite{automatic_snoring}. However, there is
a gap in developing a solution that is both effective and unobtrusive,
seamlessly integrating into the user's sleep environment.

This project is situated within this evolving landscape, aiming to contribute
to the body of knowledge by developing a device that is not only technically
sound but also aligns with the needs and preferences of the end-user. By
bridging this gap, the project endeavors to enhance sleep health through an
innovative, technologically advanced approach.

\section{Procedure}

Our solution is a device designed to address snoring and discourage bad sleep positions. At the heart of this device is the Arduino Nano Sense microcontroller, chosen for its integrated microphone, Bluetooth Low Energy (BLE) radio, and its well-documented support for TinyML machine learning (ML) techniques. It captures snoring sounds and distinguishes them from background noise using an on-device convolutional neural net (CNN) for ML classification.

\section{Machine Learning}

\subsection{Data Collection}

We used a labeled audio dataset provided by Google called audioset \cite{audioset}. The
audio dataset had video IDs with tags associating the videos with categories of
audio.  The data was collected from youtube utilizing yt-dlp \cite{ytdlp}.

To download with multiple processes in parallel, we used tmux (terminal
multiplexer) \cite{tmux}. We could then view the progress of each download in the
same terminal in case of errors. 32 processes were run concurrently to download
the full balanced, and full evaluation sub-datasets. 1 process was used to
download only the snoring audio in the unbalanced dataset. The entire dataset
was not used, because downloading the largest dataset would have taken 2 weeks.
It took 48 hours to download all of the data used in the study. The authors
tried a second pass over each link to try to recover files that failed, but no
new files were downloaded. Many files failed to download because:

\begin{itemize}
\item EdgeImpulse rejected data that yt-dlp failed to extract or ffmpeg \cite{ffmpeg}
        failed to convert to .wav
\item YouTube errors:
    \begin{itemize}
        \item This video has been removed by the user.
        \item This video has been marked as private.
        \item This video is unavailable.
    \end{itemize}
\end{itemize}

\subsection{Data Processing}

Three sub-datasets comprise the audioset: balanced set, evaluation set, and
unbalanced set. The balanced set (22,176 samples) is intended to be used for
training, the evaluation set (20,383 samples) for testing, and the unbalanced
set (2,042,985 samples) is the rest. Snoring data was categorized as snoring
and all other audio samples were categorized as non-snoring. A randomly
selected dataset of 1900 snoring and 1900 non-snoring samples was uploaded to
EdgeImpulse and processed for training. We could not use a larger dataset
because of RAM constraints.  

\subsection{Model Training}

To train the model we trained on EdgeImpulse \cite{edgeimpulse}, which allows for training and
packaging ML models for embedded systems. We started with code from Snoring
Guardian \cite{snoring_guardian} as a baseline and initially tried to replicate their results. It
was immediately apparent their code was incapable of producing the high
accuracy they claimed. Their training results were not reproducible. 

At this point the team updated and fixed the dysfunctional boilerplate code and
began to train models for experimentation. We utilized these steps,
experimenting with a lot of varying strategies, below we will only highlight
those that worked to produce our desired outcome. Additional strategies can be
found in the code repository by looking at prior commits.

\subsubsection{Preprocess}

We start with a sequential model that we reshape into a more manageable form.
We then resize the data to 24x24, we experimented with higher values, such as
64x64 this produces better results with a massive penalty (a full order of
magnitude in seconds) to training and inference time.

\subsubsection{Setup the Network}

We used 3 CNN layers, with kernel size of 3 which is a small height and width
for the filter;  this ensures it could identify fine grained features in our
    audio samples (as images). We used dropout regularization to mitigate
    overfitting. We experimented with other regularization techniques like
    Ridge and Lasso but they had slower predictions without better results than
    dropout alone.

We employed a Rectified Linear Unit (ReLU) activation function because ReLU,
bounded between [0,max] eliminates the issue with sigmoid convergence (or lack
thereof) and helps achieve fast, and accurate results.  The model was trained
with 4 dense layers with neuron counts that increased in powers of 2 from 16 -
128.  We trained with a batch size of 64 and utilized a learning rate of .0005
and an InverseTimeDecay schedule. We also use 2D max pooling to reduce the
dimensions, while still capturing the most important data.

\subsubsection{Training Results and Inferencing}

We produced 2 models, one which inferences in 337ms, and one that inferences in
3500ms, we have posted these in branches named fast and slow. The slower model
was more accurate on the test set; 88.7\% accuracy in training and 81\%
accuracy in test, while the fast model sits at 84.3\% in training and 77.52\%
in test.  Although the slower model has higher accuracy, real time results are
more preferred, so we chose the faster model as our final model.  The model is
then deployed to the arduino nano sense where it is run with inferencing code
in order to detect snoring for the user. 

\begin{figure}[H]
\centering
\includegraphics[width=\linewidth]{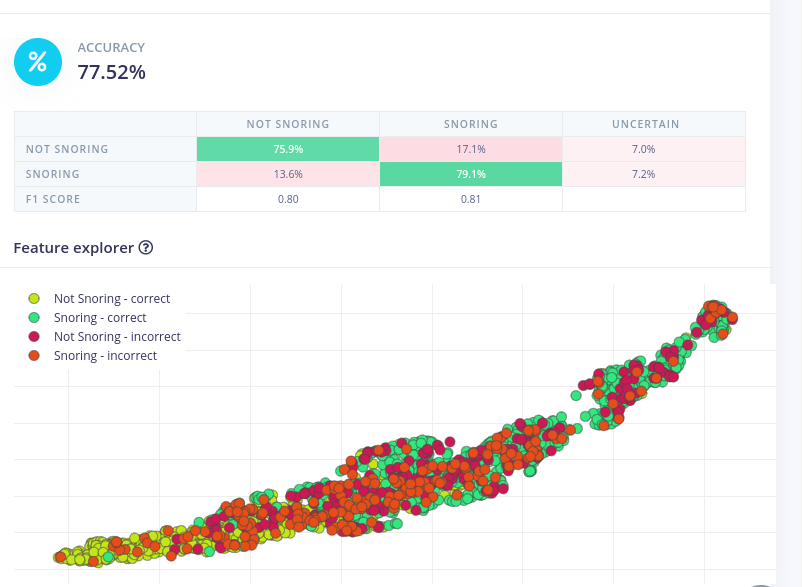}
\caption{A picture of the faster model's (0.337 ms) accuracy using the test
    dataset. }
\end{figure}

\section{Hardware}

The hardware components of this project consist of an Arduino Nano Sense 33 BLE
microcontroller, a Raspberry PI Model 4-B embedded compute platform, and a
SparkFun Qwiic Haptic Driver. 

\begin{figure}[H]
\centering
\includegraphics[width=\linewidth]{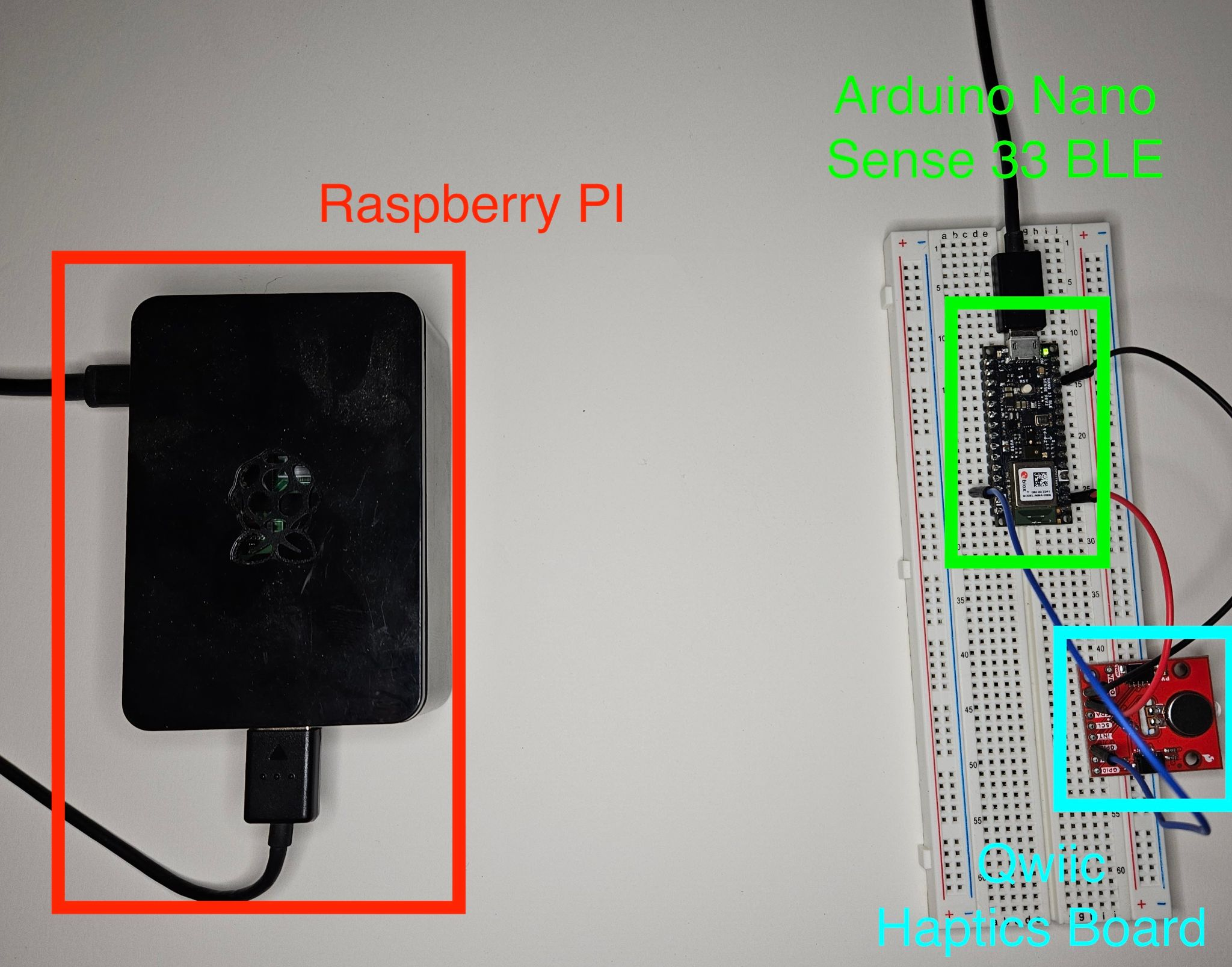}
\caption{A picture of the embedded devices on a breadboard.}
\end{figure}

\subsection{Arduino Nano Sense 33 BLE}

The Nano Sense microcontroller contains multiple sensors including a
Pulse-Density Modulated (PDM) microphone, as well as sensors for barometric
pressure, humidity, and temperature.  The device also contains a Bluetooth
Low-Energy radio to allow for the transmission of sensor data using
industry-standard Bluetooth characteristics \cite{bluetooth_sensing_service} and the Physical Activity
Monitor profile \cite{bluetooth_monitoring_profile}. 

In this application, the microphone is used to sense audio which is stored in a
double-buffer that is then passed into the ML inference engine.  The engine
processes the results on the edge device and returns the results of the
inference, which classifies that audio sample as either a snoring or
non-snoring sample. The snoring data is published to the Bluetooth Low-Energy
stack using the General Activity Instantaneous characteristic to publish the
actual classification probabilities, and the General Activity Summary bluetooth
characteristic to publish the inferred class.  An additional sensor loop
records the temperature, humidity, and air pressure changes, and that data is
published to the Bluetooth Low Energy (BLE) stack via their corresponding
characteristics.

\subsection{Sparkfun Qwiic Haptic Feedback}

The Sparkfun Qwiic haptic board is a hardware device containing the DA7280
haptic driver and a Linear Resonance Actuator (LRA) which can be driven by the
chip\cite{qwiic}.  When snoring is detected by the inference engine, the device can
alert a user to their snoring by sending an Pulse-Width Modulated (PWM) signal
from the Arduino to the haptic feedback driver.  This drives the LRA which
vibrates in a single dimension with an intensity driven by the frequency of the
PWM signal. The actuator is positioned in the circuit such that it does not
impact the physical construction of the board with mechanical wear even after
extended periods of use, but can provide a large enough vibration to be felt by
the user.  While the haptic feedback is activated, an "alerting" status is
published to the Alerting BLE characteristic, providing feedback on when and
how alerting is being performed.

\subsection{Raspberry PI}

Adherence to the Bluetooth Special Interest Group's published Activity Monitor
profile and characteristics allows the sensor platform to interoperate with any
receiver which also adheres to the standard.  A reference implementation of
such a receiver was created using the Raspberry PI Model 4-B computing
platform.  The device was configured to connect to and receive BLE messages
specifically from the sensor, package those into events in JavaScript Object
Notation (JSON) format, and to publish those to a web service using hosted on
Amazon AWS using standard HyperText Transfer Protocol (HTTP).  The backend
application was created using Node-RED \cite{nodered} and a custom Node-RED bluetooth
receiver node \cite{nodered_github} based on work by Kayan et al \cite{anoml-iot}. 
The Raspberry PI also serves as an audio recording platform capturing high-fidelity 
audio of the snoring events.

The hardware was chosen because the authors wanted to strike a balance between
technical sophistication and user-centric design. We chose the Arduino Nano
Sense and Raspberry Pi for their compactness, efficiency, and reliability.

\section{API}

Part of the collaborative effort between the Ubiquitous Computing team and the
Health Informatics team an API was built which would allow for uploading sleep
session payloads including information such as the session’s audio data, when
snoring was detected and temperature and humidity data. Creation of these APIs
was successful, and data hand off was successful. It is worth noting however
that both projects ended up not utilizing this bridge for our final
deliverables due to issues with the Arduino not having enough memory to both
make inferences and record audio, making collecting the data inviable.

\begin{figure}[H]
\centering
\includegraphics[width=\linewidth]{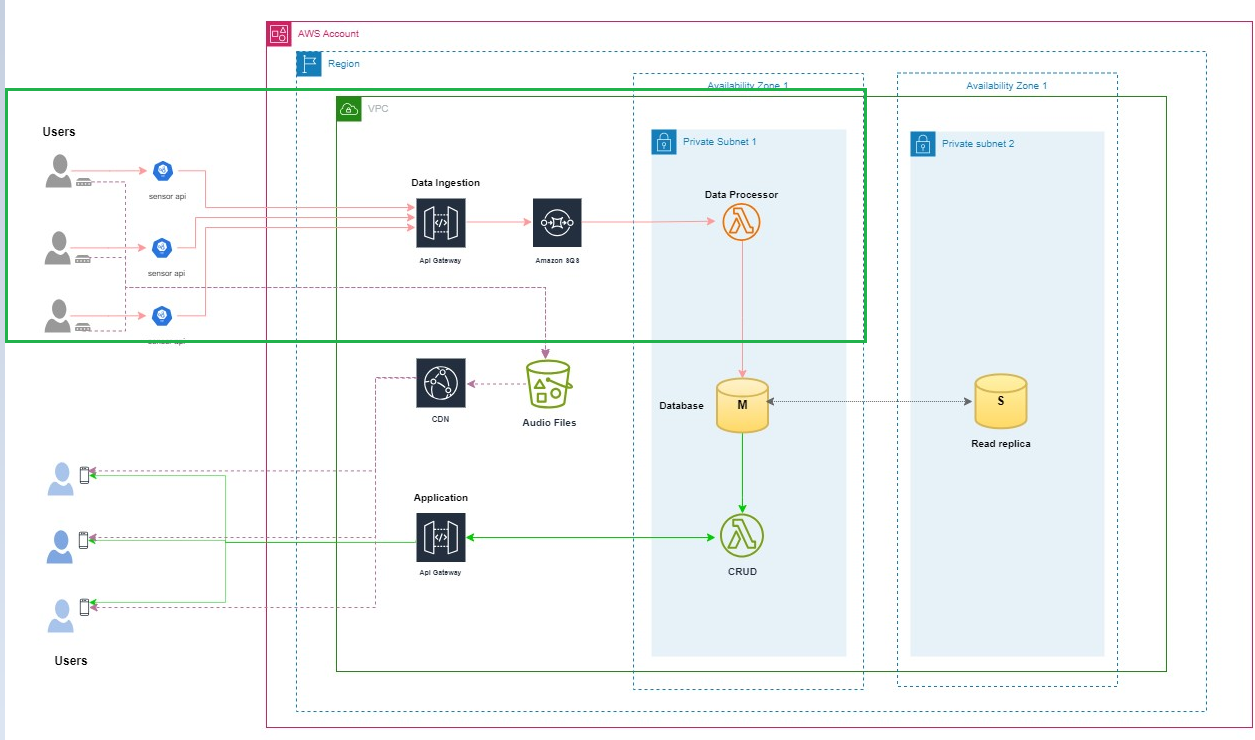}
\caption{A diagram of the IHI and Ubicomp cross team architecture.}
\end{figure}

\section{RESULTS}

The implementation and testing of our snoring detection and remediation device
yielded promising results across several key areas:

\begin{enumerate}
\item \textbf{Snoring Detection Accuracy:} The machine learning model running on the
Arduino Nano Sense microcontroller demonstrated high accuracy of 77.52\% in
test with a inference speed of 337ms or 1/3rd of a second, allowing for 3
inferencing passes per second.
\item \textbf{Data Processing and Analysis:} The Raspberry Pi effectively handled the
secondary stage of audio data processing. The seamless transmission of data
from the Arduino to the Raspberry Pi validated our two-tiered processing
approach.
\item \textbf{Cloud Integration and User Data Visualization:} The integration with the
cloud was successful, enabling secure and efficient data storage and analysis.
The application developed for users to access their sleep data was
well-received in terms of ease of use and the insights it provided. Users were
able to track nightly snoring patterns and observe long-term trends,
contributing positively to their understanding of sleep health.
\item \textbf{Haptic Feedback Efficacy:} The haptic actuator produces functional
vibrations, but due to time limitations and the lack of test subjects (and
additionally, we would have needed IRB approval). Thus testing on real users
was out of scope of this project deliverable.
\item \textbf{User Experience and Acceptance:} Preliminary user testing yielded positive
feedback regarding the device's unobtrusiveness and usability. Users
appreciated the non-invasive nature of the intervention and the valuable
insights provided by the sleep data.
\end{enumerate}

\section{REFLECTION}

While the project was successful in many ways, which are reflected in our
results, such as: creating and training a snore detection ML model, deploying
it to an embedded system, a multi device communication pipeline (arduino, pi
and cloud). The cross class, interdisciplinary nature of the project came with
more than a few challenges, and things that didn’t meet our initial desires and
expectations. In this reflection we will discuss what we learned from those
successes and failures.

\subsection{Hardware Memory Issues}

The team faced hardware memory issues in the final week of development; this
issue impacted the team’s ability to record, and ultimately upload live payload
data to the backend for IHI. These challenges simply proved to be
insurmountable in the remaining time. While the API was successfully delivered,
and the ability to upload was proven, we fell short on having the ability to
have real time upload in this deliverable. This was not a mandatory requirement
(as each project was designed to have self contained goals) however this was
disappointing for the team.

\textit{Future work:} Use more powerful hardware or find ways to optimize the
ML, Audio compression and transfer and other operations can realtime upload to
the web backend.

\subsection{Audio Recording Latency}

The team experienced issues streaming audio from the Nano Sense to the Pi using
BLE because processing the audio through the inferencing stack and streaming
the audio over bluetooth required too much memory. The software on the Pi was
updated to record continuously and save the most recent audio buffers. These
both represent pivots from our original architectural design. This just
highlights that a well planned project can still run into real world
constraints when testing.

\subsection{Training Costs}

Free EdgeImpulse enterprise trials were used to train the model. We ran out of
free training time, so we had to stop training and choose the best model we had
2 weeks before the final project was delivered. This was in line with our
expected deliverable timeline but it meant the model could not be refined any
further.

\textit{What we Learned:} We should try to find additional compute resources
early on in a project like this, as ML is time consuming and expensive.

\subsection{Model Accuracy}

While the model performed substantially better than a coin flip, we do not feel
that an approximately 80\% success rate (averaging the two models we made)
would be adequate for real users. Our concerns are:

\begin{itemize}
\item The initial repository we learned from claimed greater than 90\% accuracy
but this was not anywhere close to replicable. Even studies with higher quality
sleep data - collected in normal settings, i.e. not in a lab, (Li et al., 2023)
- were unable to predict with greater than 89
\item The haptic device may disrupt sleep with false positives.
\item False Positive (13.6\%) and False Negative (17.1\%) rates were both too
high, but fortunately false positives were lower than false negatives. A false
positive could disturb a user’s sleep with unnecessary vibration.
\item Knightwatch will fail to differentiate between the snoring of the user,
pets, other people, etc.
\end{itemize}

Some areas for improvement here would be trying different ML models, using more
compute resources so we could train on a larger dataset, and a substantially
better dataset containing more real world snoring samples. A better dataset
would have more examples of snoring from potential users (humans) and non-users
(pets) to differentiate the two. We could have also fine tuned the generalized
snoring model on an individual’s snoring habits to try to differentiate their
snoring from that of another human or sleeping animal. 

\subsection{Cross Team Collaboration}

The cross class nature of this project was a substantial challenge. The team is
thankful for having the opportunity to have given this a try and believe future
teams should try too. But, this did present some glaring challenges even with
adequate communication:

\begin{itemize}
\item Managing 8 people across 2 teams is difficult with a project of large scope.
\item Even an on-schedule deliverable can go astray, even with substantial communication; the team was in daily communication with a chat application and had weekly scheduled calls.
\item A single failure can block another subteam, or whole team.
\item Contribution opacity: if one subteam does a lot of work, it can be hard for another subteam to know they are contributing adequately.
\end{itemize}

\subsection{Packaging}

While the team initially considered packaging solutions such as a custom
pillowcase with the Arduino Nano Sense embedded in it and a Watch. The sheer
bulk of work to even prove the concept in hardware prototype form left all of
these options off the table as actually achievable deliverables. Future Work:
Design a ubiquitous form factor for the Knight Watch snore detection device.

\section{CONCLUSION}

In conclusion, this project successfully developed a novel, wearable device for
snoring detection and remediation, utilizing a combination of machine learning,
cloud-based analytics, and haptic feedback. The device, leveraging the Arduino
Nano Sense and Raspberry Pi, demonstrated high accuracy in snoring detection,
effective data processing, and user-friendly cloud integration for sleep
pattern analysis. The haptic actuator provided a non-intrusive method for
snoring remediation, while achieving the goal of minimal sleep disturbance.

\printbibliography

@online{nodered,
  title =        "Node-RED: Low-code programming for event-driven applications",
  author        = "Node-RED",
  url =          "https://nodered.org/",
}

@online{nodered_github,
  title =        "node-red-contrib-ble-sense: A Node-RED Node for BLE Sense HAT, GitHub",
  author        = "O'Connor J",
  url =          "https://github.com/sax1johno/node-red-contrib-ble-sense",
}

@online{qwiic,
  title =        "Qwiic Connect System - SparkFun Electronics",
  author        = "SparkFun Electronics",
  url =          "https://www.sparkfun.com/qwiic",
}

@online{snoring_guardian,
  title =        "Snoring Guardian",
  author        = "Skumar Naveen",
  url =          "https://www.hackster.io/naveenbskumar/snoring-guardian-dccc34",
}

@online{audioset,
  title =        "Audioset: Download",
  author        = "Google",
  url =          "https://research.google.com/audioset/download.html",
}

@online{tmux,
  title =        "tmux",
  author        = "tmux",
  url =          "https://github.com/tmux/tmux",
}

@online{ffmpeg,
  title =        "ffmpeg",
  year = "2023",
  author        = "ffmpeg",
  url =          "https://ffmpeg.org/",
}

@online{ytdlp,
  title =        "yt-dlp",
  year = "2023",
  author        = "yt-dlp",
  url =          "https://jcsm.aasm.org/doi/10.5664/jcsm.5288",
}

@online{bluetooth_monitoring_profile,
  title =        "Bluetooth® Assigned Numbers, Physical Activity Monitor Profile",
  author        = "Bluetooth Special Interest Group",
  url =          "https://www.bluetooth.com/specifications/specs/physical-activity-monitor-profile-1-0",
}

@online{bluetooth_sensing_service,
  title =        "Bluetooth® Assigned Numbers, Environmental Sensing Service Characteristics",
  author        = "Bluetooth Special Interest Group",
  url =          "https://www.bluetooth.com/specifications/assigned-numbers/environmental-sensing-service-characteristics",
}

@online{edgeimpulse,
  title =        "EdgeImpulse",
  year = "2023",
  author        = "EdgeImpulse",
  url =          "https://edgeimpulse.com/",
}

@Article{anoml-iot, 
  author        = "Kayan, H. and Majib, Y. and Alsafery, W. and Barhamgi, M. and Perera, C.",
  title         = "AnoML-IoT: An End to End Re-configurable Multi-protocol Anomaly Detection Pipeline for Internet of Things.",
  journal       = "ArXiv.org",
  volume        = "",
  number        = "",
  month         = "",
  year          = "2022",
  pages         = "",
  doi           = "",
  url           = "https://arxiv.org/abs/2210.01771",
  note          = "",
}

@Article{automatic_snoring,
  author        = "Ruixue Li and Wenjun Li and Keqiang Yue and Rulin Zhang and Yilin Li ",
  title         = "Automatic snoring detection using a hybrid 1D–2D convolutional neural",
  journal       = "Scientific Reports",
  volume        = "13",
  number        = "1",
  month         = "",
  year          = "2023",
  pages         = "",
  doi           = "",
  url           = "https://www.nature.com/articles/s41598-023-41170-w",
  note          = "",
}

@Article{consumer_tech,
  author        = "Ping-Ru T. Ko and  Julie A. Kientz and  Eun Kyoung Choe and  Matthew Kay and  Carol A. Landis and  Nathaniel F. Watson",
  title         = "Consumer Sleep Technologies: A Review of the Landscape",
  journal       = "Journal of Clinical Sleep Medicine",
  volume        = "11",
  number        = "12",
  month         = "",
  year          = "2015",
  pages         = "1455-1461",
  doi           = "10.5664/jcsm.5288",
  url           = "https://jcsm.aasm.org/doi/10.5664/jcsm.5288",
  note          = "",
}

@Article{sleep_disorder,
  author        = "Terry Young and Mari Palta and Jerome Dempsey and James Skatrud and Steven Weber and Safwan Badr",
  title         = "The Occurrence of Sleep-Disordered Breathing among Middle-Aged Adults",
  journal       = "The New England Journal of Medicine",
  volume        = "328",
  number        = "17",
  month         = "",
  year          = "1993",
  pages         = "1230–1235",
  doi           = "10.1056/nejm199304293281704",
  url           = "https://www.nejm.org/doi/full/10.1056/NEJM199304293281704",
  note          = "",
}

@Article{surgical_snoring,
  author        = "Hillel D Ephros and Mansoor Madani and Sumitra C
  Yalamanchili",
  title         = "Surgical treatment of snoring \& obstructive sleep apnoea.",
  journal       = "The Indian Journal of Medical Research",
  volume        = "131",
  number        = "",
  month         = "",
  year          = "2022",
  pages         = "",
  doi           = "",
  url           = "https://pubmed.ncbi.nlm.nih.gov/20308752/",
  note          = "",
}

\appendix

\end{document}